\documentclass[preprint, pteplogo]{ptephy_v1}
\preprintnumber{KUNS-2611, YITP-16-29} %%% %%% Insert preprint number here
\usepackage{bm,amssymb}
\usepackage{amsmath}
\usepackage{txfonts}
\usepackage{booktabs}
\usepackage{cases}

\newcommand{\Slash}[1]{{\ooalign{\hfil/\hfil\crcr$#1$}}}

\def\pbar{\overline{\psi}}

\def\Gs2{\Gamma^{(2)}_{k,\sigma}}
\def\Gp2{\Gamma^{(2)}_{k,\pi}}

\def\Jab{J_{k,\alpha\beta}}
\def\Jss{J_{k,\sigma\sigma}}
\def\Jpp{J_{k,\pi\pi}}
\def\Jsp{J_{k,\sigma\pi}}
\def\Jps{J_{k,\pi\sigma}}

\def\Is{I^{(2)}_{k,\sigma}}
\def\Ip{I^{(2)}_{k,\pi}}

\def\Jsqq{J^{(\sigma)}_{k,\bar{\psi}\psi}}
\def\Jpqq{J^{(\pi)}_{k,\pbar\psi}}

\def\Dk{\partial_{k}}

\def\RB{R^{B}_{k}}
\def\RF{R^{F}_{k}}

\def\Ea{E_{\alpha}}
\def\Eb{E_{\beta}}
\def\Eq{E_{\psi}}

\def\Ebt{\tilde{E}_{\beta}}
\def\Eqt{\tilde{E}_{\psi}}

\def\ip0{\mathrm{i}p_{0}}
\def\p02{p_{0}^{2}}

\def\q2{\vec{q}^2}
\def\qabs{|\vec{q}|}

\newcommand{\MeV}{\mathrm{MeV}}
\newcommand{\pab}{
\ifmmode p
\else $p$
\fi
}
\newcommand{\pfour}{
\ifmmode P
\else $P$
\fi
}
\newcommand{\qfour}{
\ifmmode Q
\else $Q$
\fi
}
\allowdisplaybreaks[1]

\begin{document}

\title{Functional renormalization group analysis of the soft mode at the QCD critical point}

\author{Takeru Yokota$^{1,\ast}$, Teiji Kunihiro$^{1}$, and Kenji Morita$^{2}$}

\affil{$^1$Department of Physics, Faculty of Science, Kyoto University, Kyoto 606-8502, Japan\\
 $^2$Yukawa Institute for Theoretical Physics, Kyoto University, Kyoto 606-8502, Japan
\email{tyokota@ruby.scphys.kyoto-u.ac.jp}} 

\begin{abstract}
 We make an intensive investigation of the soft mode at the quantum chromodynamics (QCD) critical
 point on the basis of the functional renormalization group (FRG)
 method in the local potential approximation. We calculate
the spectral functions
 $\rho_{\sigma, \pi}(\omega,\, \pab)$ in the scalar ($\sigma$) and
 pseudoscalar ($\pi$)  channels beyond the random phase approximation in
 the quark--meson model. At finite baryon chemical potential $\mu$ with a
 finite quark mass, the baryon-number fluctuation is coupled to the
 scalar channel and the spectral function in the $\sigma$ channel has a
 support not only in the time-like ($\omega\,>\,p$) but also in the
 space-like ($\omega\,<\, \pab$) regions, which correspond to the
 mesonic and the particle--hole phonon excitations, respectively. We find
 that the energy of the peak position of the latter becomes vanishingly
 small with the height being enhanced as the system approaches the
 QCD critical point, which is a manifestation of the fact that the
 phonon mode is the {\em soft mode} associated with the second-order
 transition at  the QCD critical point, as has been suggested by some
 authors.  Moreover, our extensive calculation of the spectral function
 in the $(\omega, \pab)$ plane enables us to see that the mesonic and phonon modes have the respective definite dispersion relations
 $\omega_{\sigma.{\rm ph}}(\pab)$, and it turns out that
 $\omega_{\sigma}(\pab)$ crosses the light-cone line into the
 space-like region, and then eventually merges into the phonon mode as
 the system approaches the critical point more closely. This implies
 that the sigma-mesonic mode also becomes soft at the critical point.
 We also provide numerical stability conditions that are necessary for obtaining
 the accurate effective potential from the flow equation.
\end{abstract}

\subjectindex{B32, D30, D31}

\maketitle

\section{Introduction}
\label{Intro}
The phase diagram of quantum chromodynamics (QCD) is expected to have
a rich structure and its clarification is one of the main topics in
high-energy and nuclear physics \cite{Fukushima:2010bq}.
One of the remarkable features in the expected phase structure is the
existence of the first-order phase boundary between the hadronic phase
and the quark--gluon plasma (QGP)  phase at large baryon chemical potential {$\mu$}. 
In particular, the end point of the first-order phase boundary is known
as the QCD critical point, where the phase transition becomes second
order. Although lattice QCD, which is a powerful nonperturbative
first-principle method for QCD, has a limited predictive power in the
case of large {$\mu$}  because of the sign problem
\cite{Muroya:2003,deForcrand:2009, Aarts:2015},
other QCD-motivated approaches, such as chiral effective models
implementing relevant symmetries and functional methods with inputs from
lattice QCD, support the existence of the QCD critical point
%NJL, PNJL, QM, PQM, RM and Schwinger-Dyson
\cite{Asakawa:1989bq,Hatsuda:1994pi, Halasz:1998,Scavenius:2001,Buballa:2003qv,
Fukushima:2003fw, Ratti:2005jh, Schaefer:2005, Schaefer:2007pw,
Herbst:2013, Fischer:2014, Stiele:2016cfs}.

While the location of the QCD critical point in these calculations
strongly depends on details of the models and employed approximations
\cite{Stephanov:2004PTEP,Sasaki:2007, Kashiwa:2008, Nakano:2010}, 
one may expect anomalous fluctuations of experimental observables 
in relativistic heavy ion collisions if the produced matter passes through
critical region where the thermodynamic quantities are strongly
influenced by the existence of the critical point
\cite{Stephanov:1999zu}. In particular, the net-baryon number
susceptibility and its higher-order ones, approximated by the net-proton
ones in measurements \cite{Hatta-Stephanov}, are expected to be sensitive to the critical
behavior of the system \cite{Hatta:2002sj, Sasaki:2007, Asakawa:2009prl,
Stephanov:2009prl, Stephanov:2011prl, Skokov:2011prc, Friman:2011epjc,
Morita:2013prob, Morita:2013prob2, Morita:2013prob3, Ichihara:2015}. 
A recent summary of measurements in the beam-energy scan program at the Relativistic
Heavy Ion Collider (RHIC) including intriguing behaviors of the
net-proton fluctuations can be found in Ref. \cite{BES}.

Although the mean-field theory seems to give a reasonable picture
of the phase structure, a system near a critical point exhibits strong
correlations among its constituents. Therefore methods beyond the
mean-field theory are desirable for understanding the physics near the
critical point. The functional renormalization group (FRG)
\cite{Wetterich:1992yh,Wegner:1972ih,Wilson:1974,Polchinski:1983gv}
is one of the frameworks that enables us to incorporate
fluctuation effects beyond the mean-field theory; see also
 Refs. \cite{Berges:2000ew,Pawlowski:2005xe,Gies:2006wv}.
Indeed, it has been pointed out that such fluctuations play an important role in the aforementioned
observables \cite{Skokov:2010,Skokov:2011prc, Morita:2013prob, Morita:2013prob2, Morita:2013prob3,
Ichihara:2015, Morita:2015,Fu-Pawlowski:2015}. 
The FRG has been applied to a wide range of fields including hot and dense QCD and found to be useful in the description of chiral phase
transition in QCD using effective chiral models
\cite{Jungnickel:1995fp,Braun:2003ii,Schaefer:2005,Schaefer:2006ds,
Stokic:2010, Nakano:2010, Aoki:2014}.

For a second-order transition, there exist specific modes that are coupled to the 
fluctuations of the order parameter and become gapless
and long-lived at
the critical point. Such a mode is called the soft mode of the phase transition.
As for the QCD critical point, the nature of the soft modes changes
depending on whether the quarks have a mass or not
\cite{Fujii:2004jt,Son:2004iv}:
In the chiral limit where the quarks are massless, the theory has exact
chiral symmetry and an O(4) {\em critical line} 
appears in the $(T,\,\mu)$ plane in the two-flavor case; the critical
line terminates at a point (the tricritical point) in the
$(T,\,\mu)$ plane and is connected to a first-order phase transition
line \cite{Pisarski:1984}; the $\sigma$ and $\pi$ mesonic modes in the
time-like region become massless on the critical line, implying that the
quartet mesons are the soft modes of the chiral transition in the
chiral limit \cite{Hatsuda:1985ey,Hatsuda:1985ey2,Hatsuda:1985eb}.
At finite baryon chemical potential, however, the picture changes
because the charge conjugation symmetry is lost. 
When the chiral symmetry is broken, the mixed correlator 
$\langle\langle \bar{\psi}\psi\bar{\psi}\gamma^0\psi\rangle\rangle$ 
does not vanish. Then, when the system approaches the tricritical point in the
broken phase, the density fluctuation  
$\langle\langle (\bar{\psi}\gamma^0\psi)^2\rangle\rangle$ 
also shows a singular behavior.
Furthermore, when the theory does not have chiral symmetry due to the
current quark masses from the outset, the natures of the phase transition
and the soft mode change drastically. In this case, the tricritical
point becomes the critical point where the universality class 
belongs to that of the $Z_{2}$ critical point and the soft mode is
considered to be the particle--hole mode corresponding to the density
(and energy) fluctuations. It is noteworthy that not only the chiral susceptibility but also the susceptibilities of hydrodynamical modes such as the density
fluctuation or the quark-number susceptibility diverge at the critical point,
owing to the scalar--vector coupling caused by the finite quark mass at
nonvanishing $\mu$ \cite{Kunhiro:1991}.
Such a picture is suggested in the time-dependent Ginzburg--Landau theory and
random phase approximation (RPA) analysis of the Nambu--Jona-Lasino (NJL)
model in Ref. \cite{Fujii:2004jt} and the Langevin equation in Ref. \cite{Son:2004iv}.

Then it would be intriguing to apply the FRG for investigating the 
nature and dynamical properties of the soft mode at the critical point, which is
the purpose of the present work. Here it should be mentioned that
the role of density fluctuations in the static critical properties of the QCD critical point
has been investigated in FRG within a chiral quark--meson model
\cite{Kamikado:2012cp}, where static properties such as 
the quark--number susceptibility and the curvature (screening) masses 
in the scalar and vector channels are analyzed. Dynamical properties such as the 
dispersion relations of excitation modes and those of the soft modes
at the QCD critical point have not been touched upon. We note that
these quantities can be read off from the spectral function in the relevant channel.

Needless to say, a real-time analysis  is needed for the investigation
of the spectral functions for excitation modes. Since analytic
continuation of the two-point functions from imaginary Matsubara frequencies
to real frequencies is necessary to get real-time two-point Green's
functions at finite temperature, it is in general difficult to obtain the
spectral functions numerically with high accuracy
\cite{Jarrell:1996,Asakawa:2000tr,Vidberg:1977,Dudal:2013yva}. 
Recently, a useful method for calculating the spectral functions within FRG
has been developed, which adopts an unambiguous way of analytic
continuation in the imaginary-time formalism and leads to reasonable
results of meson spectral functions in the O(4) model in vacuum
\cite{Kamikado:2013sia}. Furthermore, this method has also been successfully applied to the quark--meson model at finite temperature and
chemical potential \cite{Tripolt:2013jra,Tripolt:2014wra}.

The spectral function in a specific channel with specific quantum numbers
may have more than one peak and bump, the number of which can change
according to that of the parameters characterizing the system such as
temperature and baryon density. A peak or even bump in the
spectral function in the channel may be identified with a particle
excitation in the channel,  and the width of the peak/bump shows the
decay rate of the particle in a specific decay channel.
This feature enables one not only to explore the appearance but also to
analyze the nature of modes in the system using spectral functions in
various channels. In particular, the spectral functions around a
critical point give information on the soft modes. Actually, the
spectral function in the scalar--isoscalar channel, i.e., the sigma channel, has been analyzed in the RPA using the Nambu--Jona-Lasinio (NJL) model in Ref. \cite{Fujii:2004jt},
where it is shown that the spectral function has a prominent bump 
in the space-like region corresponding to the phonon mode composed of 
particle--hole excitations, the peak position of which moves to
vanishingly small frequency as the system approaches the critical point,
whereas the mesonic sigma mode 
in the time-like region does not show such a behavior, retaining a finite
mass. This result clearly shows that the phonon mode (or hydrodynamical
mode in general) is the soft mode of the QCD critical point; see also Refs. \cite{Son:2004iv,Minami:2009hn,Minami:2009hn2,Minami:2009hn3}. 

The purpose of the present paper is to investigate the nature of
low-energy modes at the QCD critical point beyond the RPA using a two-flavor
quark--meson model:
We calculate the spectral functions in the $\sigma$ 
and pion channels with FRG.
Our results confirm the softening of the particle--hole mode 
in the $\sigma$ channel near the QCD
critical point, but not in the pion channel. 
In addition, we find that the low-momentum dispersion
relation of the sigma-mesonic mode penetrates into the space-like region and the
mode merges into the bump of the particle--hole mode.

This paper is organized as follows.
In Sect. \ref{SMethod}, we recapitulate the method 
\cite{Tripolt:2013jra,Tripolt:2014wra} and
describe how to calculate the
spectral functions in the mesonic channels numerically
with FRG.
The numerical results are shown in
Sect. \ref{SNumericalResults}.
The phase diagram, the critical region, and the precise location of the
critical point are presented in Sect. \ref{SSPhaseDiagram}.
In Sect. \ref{SSSpectral}
the results of the spectral functions are
shown, and  the soft mode at the QCD critical point is discussed. Section \ref{SSummary} is devoted to the summary and outlook.
In Appendix \ref{AppFlow}, the detailed forms
of the FRG flow equations after Matsubara summation 
are shown. In Appendix \ref{AppStability},
we derive the conditions for numerically stable calculation of
the flow equation as a nonlinear partial differential equation.

\section{Method} \label{SMethod}

In this section, we recapitulate the method developed in Refs. \cite{Tripolt:2013jra,Tripolt:2014wra} for calculating spectral
functions in the quark--meson model with FRG,
 and present some details of
our numerical procedure.

\subsection{Procedure to derive spectral functions in meson channels}

\begin{figure}[!b]
\centering\includegraphics[width=0.9\columnwidth]{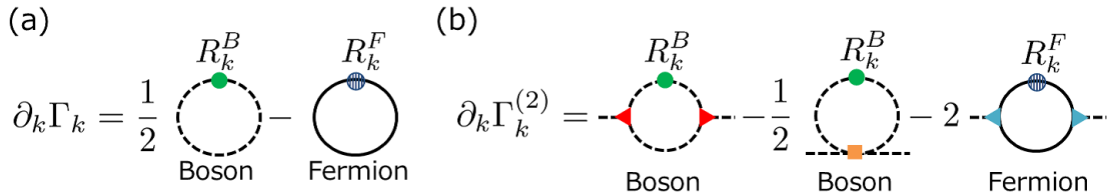}
\caption{Diagrammatic representations of (a) Eq. (\ref{Weq})
and (b) Eq. (\ref{G2flow}).}
\label{DiagRep}
\end{figure}

The FRG is based on the philosophy of the Wilsonian renormalization group
and realizes the coarse graining by introducing a regulator function
$R_{k}$ , which has the role of suppressing lower-momentum modes than the
scale $k$ for the respective field. In this method, the effective
average action (EAA) $\Gamma_{k}$ is introduced such that it becomes bare action
$S_{\Lambda}$ at a large UV scale $k=\Lambda$ and becomes the effective
action at $k\rightarrow 0$ with an appropriate choice of
regulators. The flow equation for EAA, the Wetterich equation,
can be derived as a functional differential equation \cite{Wetterich:1992yh}:
\begin{equation}
 \partial_{k}\Gamma_{k}=\frac{1}{2}\mathrm{STr}\left[
\frac{\partial_{k}R_{k}}{\Gamma^{(2)}_{k}+R_{k}}
\right],
\label{Weq}
\end{equation}
where $\Gamma_{k}^{(n)}$ is the $n$th functional derivative of
$\Gamma_{k}$ with respect to fields. This equation has a one-loop structure
and can be represented diagrammatically (Fig. \ref{DiagRep} (a)).

In principle, one can get the effective action $\Gamma_{k=0}$ by solving Eq. (\ref{Weq}) 
with the initial condition $\Gamma_{\Lambda}=S_{\Lambda}$.

However, it is prohibitively difficult to solve Eq. (\ref{Weq}) 
in an exact way and some simplifications are introduced for practical use. One of the simplifications is to truncate the form of EAA. Since our purpose is to reveal the behavior of the low-momentum modes
around the QCD critical point, we adopt a truncation in which the
low-momentum fluctuations are properly taken into account. 

We should now specify the low-energy effective model of QCD; we employ
the two-flavor quark--meson model as such a model. Then we take the local
potential approximation (LPA) for the meson flow part as our
truncation. This truncation corresponds to considering only
the lowest order of derivative expansion for the meson flow part.

In the imaginary-time formalism, our truncated EAA is as follows
\cite{Schaefer:2006ds}:
\begin{align}
\Gamma_{k}\left[\pbar,\psi,\phi \right]
=&
\int_{0}^{\frac{1}{T}}d\tau
\int d^{3}x
\left\lbrace
\pbar
\left(\Slash{\partial}
+g_{s}(\sigma+i\vec{\tau}\cdot\vec{\pi}\gamma_{5})
-\mu \gamma_{0} \right)
\psi
+\frac{1}{2}(\partial_{\mu}\phi)^2
+U_{k}(\phi^2)-c\sigma
\right
\rbrace,
\label{QMtranc}
\end{align}
where $\phi=(\sigma,\vec{\pi})$. The quark field $\psi$ has the indices
of a four-component spinor, color $N_{c}=3$, and flavor $N_{f}=2$.
The $c\sigma$ represents the effect of the current quark mass, which
explicitly breaks $N_{f}=2$ chiral symmetry. In this truncation, we also
neglect the flow of $g_s$ and the wave function renormalization. Therefore,
only the meson effective potential $U_{k}$ has a $k$ dependence.

Although the truncated EAA itself gives only the simplest information about
the momentum dependence of two-point Green's functions, 
the nonperturbative effects are to be incorporated through Eq. (\ref{Weq})
with the truncated EAA used as the initial condition. More specifically,
we first calculate the effective potential $U_{k}(\phi^{2})$ using
Eq. (\ref{Weq}). Then the chiral condensate $\sigma_{0}$ is obtained 
as $\sigma$ satisfying the quantum equation of motion
(EOM) $\delta \Gamma_{k=0}/\delta \sigma = 0$. In our case,
this condition corresponds to obtaining $\sigma$ that minimizes
$U_{k}(\sigma^{2})-c\sigma$,
under the assumptions that the condensate is homogeneous
and $\langle \vec{\pi} \rangle = \vec{0}$. Next, we derive the flow
equations for two-point Green's functions. Such  flow equations 
can be derived from the flow equation for $\Gamma^{(2)}_{k}$
by differentiating both sides of Eq. (\ref{Weq}):
\begin{align}
\partial_{k}\Gamma^{(2)}_{k}=&\mathrm{STr}
\left[
\frac{1}{\Gamma^{(2)}_{k}+R_{k}}
\Gamma^{(3)}_{k}
\frac{1}{\Gamma^{(2)}_{k}+R_{k}}
\Gamma^{(3)}_{k}
\frac{1}{\Gamma^{(2)}_{k}+R_{k}}
\partial_{k}R_{k}
\right]
-\frac{1}{2}
\mathrm{STr}
\left[
\frac{1}{\Gamma^{(2)}_{k}+R_{k}}
\Gamma^{(4)}_{k}
\frac{1}{\Gamma^{(2)}_{k}+R_{k}}
\partial_{k}R_{k}
\right].
\label{G2flow}
\end{align}
The diagrammatic expression of this equation is shown in Fig. \ref{DiagRep} (b).
If the fields in Eq. (\ref{G2flow}) are replaced by those satisfying
the quantum EOM, Eq. (\ref{G2flow}) becomes the flow equations for the
inverses of the two-point Green's functions.
Then, Eq. (\ref{G2flow}) can be rewritten as the flow equations for the
inverses of the two-point temperature Green's functions $\Gamma^{(2)}_{k,\sigma}(\pfour)$ 
and $\Gamma^{(2)}_{k,\pi}(\pfour)$ in the sigma
 and pion channels, respectively, which are defined  as
\begin{align}
\left.
\frac{\delta^{2} \Gamma_{k}}{\delta \sigma(\pfour) \delta \sigma(\qfour)}
\right|
_{\pbar=0,\psi=0,\vec{\pi}=\vec{0},\sigma=\sigma_{0}}
&=(2\pi)^{4}\delta^{(4)}(\pfour + \qfour)\Gamma^{(2)}_{k,\sigma}(\pfour), \\
\left. 
\frac{\delta^{2} \Gamma_{k}}{\delta \pi_{a}(\pfour) \delta \pi_{a}(\qfour)}
\right|
_{\pbar=0,\psi=0,\vec{\pi}=\vec{0},\sigma=\sigma_{0}}
&=(2\pi)^{4}\delta^{(4)}(\pfour + \qfour)\Gamma^{(2)}_{k,\pi}(\pfour),
\end{align}
where $\sigma(P)$ and $\pi_{a}(P)$ are the Fourier transforms of
$\sigma (x)$ and $\pi_{a}(x)$, respectively, and $\pfour=(i\omega_{n},\vec{p})$ with
$\omega_{n}$ being the bosonic Matsubara frequency.
The RHS of Eq. (\ref{G2flow}) contains $\Gamma_{k}^{(2)}$,
$\Gamma_{k}^{(3)}$, and $\Gamma_{k}^{(4)}$.
In general, the flow equation comprises an infinite hierarchy of differential equations 
such that the flow equation for $\Gamma_{k}^{(n)}$ contains
$\Gamma_{k}^{(n+1)}$ and $\Gamma_{k}^{(n+2)}$. We can simplify the flow equation for the $\Gamma^{(n)}_{k}$
by replacing these derivatives with those derived from the truncated
EAA. Under the above procedure, the integration of the flow equation for
$\Gamma^{(2)}_{k,\sigma(\pi)}(\pfour)$ leads to the two-point Green's
function in the sigma (pion) channel with the nonperturbative effects incorporated. 

A real-time two-point Green's function at finite temperature is obtained
by analytic continuation for  the temperature Green's function to real
time, i.e., from imaginary Matsubara frequencies to real frequencies in
the case of momentum representation.
In our case, such an analytic continuation is successfully carried
out at the level of the flow equation after the Matsubara summations in
the RHS of Eq. (\ref{G2flow}), as follows:
When one derives a retarded two-point Green's function
via analytic continuation in the frequency $\omega$ plane,
the analyticity of the Green's function in the upper half-plane of $\omega$ 
must be retained \cite{BM1961}.
In the present case, the flow equation itself should be analytic
in the upper half-plane after the analytic continuation.
One can retain the analyticity in the upper half-plane
easily by taking into account the following points.
First, by choosing $\omega_n$-independent regulators, 
one can
avoid possible extra poles in the $\omega$ plane in the flow equation
 otherwise arising from $\omega_n$ dependence of the regulators.
The second point is about the analytic continuation of thermal
distribution functions $n_{B,F}(E+i\omega_{n})$ obtained for a discrete (multiple of $2\pi T$)
frequency $\omega_n$, where
the subscript $B$, $F$ stands for a boson or fermion, respectively,  and $E$ is $\omega_{n}$
independent.
Such factors appear in the flow equation after the
Matsubara summation.
Because of the periodicity of the exponential function,
$n_{B,F}(E+i\omega_{n})$ is equal to $n_{B,F}(E)$.
However if $n_{B,F}(E+\omega)$
is substituted for $n_{B,F}(E+i\omega_{n})$,
such a factor breaks the analyticity of the flow equation
in the upper half-plane.
Therefore $n_{B,F}(E+i\omega_{n})$ should be
replaced by $n_{B,F}(E)$ before the analytic continuation.
By taking into account these points,
the substitution $\omega+i\epsilon$ for $i\omega_{n}$
with $\epsilon$ being a positive infinitesimal
 gives 
the flow equation for the inverse of the retarded Green's function
$\Gamma^{(2),R}_{k,\sigma,\pi}(\omega,\vec{p})$.
Finally, the spectral functions in the meson channels are given in terms of the thus-obtained retarded Green's functions as follows:
\begin{align}
\rho_{\sigma}(\omega, \vec{p})&=\frac{1}{\pi}\mathrm{Im}
\frac{1}{\Gamma^{(2),R}_{k\rightarrow 0,\sigma}(\omega,\vec{p})}, \\
\rho_{\pi}(\omega, \vec{p})&=\frac{1}{\pi}\mathrm{Im}
\frac{1}{\Gamma^{(2),R}_{k\rightarrow 0,\pi}(\omega,\vec{p})}.
\end{align}

\subsection{Flow equations}
In the present work, we adopt the 3D Litim's optimized
regulators for bosons and fermions \cite{Litim:2001up}
as $\omega_{n}$-independent regulators:
\begin{eqnarray}
\RB(\qfour)&=&(k^2-\vec{q}^2)\theta(k^2-\vec{q}^2), \\
\RF(\qfour)&=&i\Slash{\vec{q}}\left(\sqrt{\frac{k^2}{\vec{q}^2}}-1\right)\theta(k^2-\vec{q}^2).
\end{eqnarray}
Then the insertion  of Eq. (\ref{QMtranc})
into Eq. (\ref{Weq}) leads to the following flow equation for $U_{k}$:
\begin{align}
\partial_{t}U_k
=\frac{k^5}{12\pi^2}
&\left[
-2N_f N_c
\left[
\frac{1}{E_{\psi}}
\tanh\frac{E_{\psi}+\mu}{2T}
+
\frac{1}{E_{\psi}}
\tanh\frac{E_{\psi}-\mu}{2T}
\right]
+
\frac{1}{E_{\sigma}}
\coth\frac{E_{\sigma}}{2T}
+
\frac{3}{E_\pi}
\coth\frac{E_\pi}{2T}
\right],
\label{Ukflow}
\end{align}
where $t=\ln(k/\Lambda)$, $E_{a}=\sqrt{k^2+m^2_{a}}\ (a=\psi ,\sigma,\pi)$, and
\begin{align}
m^2_{\psi}=g^2_s \sigma^2,\  m^2_{\sigma} =\partial^2_{\sigma}U_k,\  m^2_{\pi} =\partial_\sigma U_k /\sigma .
\label{Masses}
\end{align}

According to the procedure presented in the previous subsection,
the flow equations for $\Gamma^{(2)}_{k,\sigma}(\pfour)$ and
$\Gamma^{(2)}_{k,\pi}(\pfour)$ become:
\begin{align}
\Dk\Gs2(\pfour)
=&
\Jss(\pfour)(\Gamma^{(0,3)}_{k,\sigma\sigma\sigma})^{2}
-\frac{1}{2}\Is \Gamma^{(0,4)}_{k,\sigma\sigma\sigma\sigma}
+3\Jpp(\pfour)(\Gamma^{(0,3)}_{k,\sigma\pi\pi})^{2}
-\frac{3}{2}\Ip \Gamma^{(0,4)}_{k,\sigma\sigma\pi\pi} \notag
\\
&-2N_{c}N_{f}\Jsqq(\pfour),
\label{G2Sflow} \\
\Dk\Gp2(\pfour)
=&
\Jsp(\pfour)(\Gamma^{(0,3)}_{k,\sigma\pi\pi})^{2}+\Jps(\pfour)(\Gamma^{(0,3)}_{k,\sigma\pi\pi})^{2}
-\frac{1}{2}\Is\Gamma^{(0,4)}_{k,\sigma\sigma\pi\pi}
-\frac{5}{2}\Ip\Gamma^{(0,4)}_{k,\pi\pi\tilde{\pi}\tilde{\pi}} \notag \\
&-2N_{c}N_{f}\Jpqq(\pfour),
\label{G2Pflow}
\end{align}
respectively, where $\pi,\tilde{\pi} \in \lbrace \pi_{1},\pi_{2},\pi_{3}\rbrace$ and $\pi\neq \tilde{\pi}$.
The loop-functions
$J_{k,\alpha\beta}(P)$,\, $I^{(2)}_{k,\alpha}$, and $J_{k,\pbar \psi}^{(\alpha)}(P)$\, $(\alpha,\beta=\sigma,\pi)$ are defined as
\begin{align}
J_{k,\alpha\beta}(\pfour)&=T\sum_{q_{n}}\int \frac{d^{3}\vec{q}}{(2\pi)^{3}} \partial_{k} R_{k}^{B}(q)
G^{B}_{k,\alpha}(P)^{2} G^{B}_{k,\beta}(Q-P),
\label{Jab}
\\
I^{(2)}_{k,\alpha}&=T\sum_{q_{n}}\int \frac{d^{3}\vec{q}}{(2\pi)^{3}} \partial_{k} R_{k}^{B}(q)
G^{B}_{k,\alpha}(Q)^{2},
\label{Ia}
\\
J_{k,\pbar \psi}^{(\alpha)}(\pfour)&=T\sum_{q_{n}}\int \frac{d^{3}\vec{q}}{(2\pi)^{3}} \mathrm{tr}
\left[
\Gamma^{(2,1)}_{\pbar\psi\alpha}
G^{F}_{k,\pbar \psi}(Q)
\partial_{k}R_{k}^{F}(Q)
G^{F}_{k,\pbar \psi}(Q)
\Gamma^{(2,1)}_{\pbar\psi\alpha}
G^{F}_{k,\pbar \psi}(Q-P)
\right],
\label{Jqq}
\end{align}
where $\qfour=(iq_{n},\vec{q})$ and
\begin{align}
G^{B}_{k,\alpha}(\qfour)&=\left[\qfour^{2}+\left. m_{\alpha}^{2}\right|_{\sigma=\sigma_{0}}+R_{k}^{B}(\qfour)
\right]^{-1},
\\
G^{F}_{k,\pbar \psi}(\qfour)&=
\left[\Slash{\qfour}-\mu \gamma_{0}+\left. m_{\psi}\right|_{\sigma=\sigma_{0}}+R^{F}_{k}(\qfour)\right]^{-1}.
\end{align}
The three- and four-point vertices $\Gamma^{(2,1)}_{\pbar\psi\phi_{i}}$,\,
$\Gamma^{(0,3)}_{k,\phi_{i}\phi_{j}\phi_{l}}$,
and $\Gamma^{(0,4)}_{k,\phi_{i}\phi_{j}\phi_{l}\phi_{m}}$
are defined as
\begin{align}
\frac{\delta}{\delta \phi_{i} (P_{1})}
\frac{\overset{\rightarrow}{\delta}}{\delta \pbar (P_{2})}
\Gamma_{k}
\frac{\overset{\leftarrow}{\delta}}{\delta \psi (P_{3})}
&=(2\pi)^{4}\delta^{(4)}(P_{1}+P_{2}+P_{3})
\Gamma^{(2,1)}_{\pbar\psi\phi_{i}},
\\
\frac{\delta^{3} \Gamma_{k}}{
\delta \phi_{i} (P_{1})
\delta \phi_{j} (P_{2})
\delta \phi_{l} (P_{3})}
&=(2\pi)^{4}\delta^{(4)}(P_{1}+P_{2}+P_{3})
\Gamma^{(0,3)}_{k,\phi_{i}\phi_{j}\phi_{l}}, \\
\frac{\delta^{4} \Gamma_{k}}{
\delta \phi_{i} (P_{1})
\delta \phi_{j} (P_{2})
\delta \phi_{l} (P_{3})
\delta \phi_{m} (P_{4})}
&=(2\pi)^{4}\delta^{(4)}(P_{1}+P_{2}+P_{3}+P_{4})
\Gamma^{(0,4)}_{k,\phi_{i}\phi_{j}\phi_{l}\phi_{m}},
\end{align}
some of which are expressed in terms of $U_{k}$:
\begin{align}
\Gamma^{(2,1)}_{\pbar\psi\phi_{i}}
&=
\begin{cases}
g_{s} &(\text{for }i=0)\\
g_{s}i\gamma^{5}\tau^{i} &(\text{for }i=1,2,3)
\end{cases},
\\
\Gamma^{(0,3)}_{k,\phi_{i}\phi_{j}\phi_{l}}
&=4U_{k}^{(2)}
(\delta_{ij}\phi_{m}+\delta_{im}\phi_{j}+\delta_{jm}\phi_{i})
+8U_{k}^{(3)}\phi_{i}\phi_{j}\phi_{m},
\\
\Gamma^{(0,4)}_{k,\phi_{i}\phi_{j}\phi_{l}\phi_{m}}
&=4U_{k}^{(3)}
(\delta_{ij}\delta_{mn}+\delta_{in}\delta_{jm}+\delta_{jn}\delta_{im}) \notag \\
&+8U_{k}^{(3)}(
\delta_{ij}\phi_{l}\phi_{m}
+\delta_{jl}\phi_{i}\phi_{m}
+\delta_{lm}\phi_{i}\phi_{j}
+\delta_{jm}\phi_{i}\phi_{l}
+\delta_{im}\phi_{j}\phi_{l}
+\delta_{il}\phi_{j}\phi_{m}
) \notag \\
&+16U_{k}^{(4)}\phi_{i}\phi_{j}\phi_{l}\phi_{m}.
\end{align}
Analytic continuation in Eq. (\ref{G2Sflow}) and Eq. (\ref{G2Pflow})
is carried out after the Matsubara summation in Eqs. (\ref{Jab})--(\ref{Jqq}).
The detailed forms of Eqs. (\ref{Jab})--(\ref{Jqq}) after
Matsubara summation are presented in Appendix \ref{AppFlow}.

To solve these flow equations, 
we employ the following initial conditions at the UV scale $k=\Lambda$:
\begin{align}
U_{\Lambda}(\phi^2)&=\frac{1}{2}m_{\Lambda}^2\phi^2+\frac{1}{4}\lambda_{\Lambda}(\phi^2)^2, \label{initial}
\\
\Gamma_{\Lambda ,\sigma}^{(2),R}(\omega,\vec{p})
&=-\omega^{2}+\vec{p}^{2}+\partial^2_{\sigma}U_{\Lambda}(\sigma_{0}^{2}),  \\
\Gamma_{\Lambda ,\pi}^{(2),R}(\omega,\vec{p})
&=-\omega^{2}+\vec{p}^{2}+\partial_\sigma U_{\Lambda}(\sigma_{0}^{2}) /\sigma_{0}.
\end{align}

\subsection{Numerical procedure}
We employ the grid method to solve Eq. (\ref{Ukflow}) numerically.
This method reveals the global structure of $U_{k}(\sigma^{2})$ on
discretized $\sigma$. We employ the fourth-order Runge--Kutta method to solve Eq. (\ref{Ukflow}).
It is known that certain conditions between the intervals
of discretization of variables must be fulfilled to solve
 partial differential equations numerically in a stable way \cite{NumericalRecipes}.
We have derived such conditions for Eq. (\ref{Ukflow}), which are presented in Appendix \ref{AppStability}.
We fix the intervals of discretization of $\sigma$ and $t$ 
in Eq. (\ref{Ukflow})
according to these conditions.
As stated before, the flow equation (\ref{Ukflow}) should be
solved down to $k=0$ from $k=\Lambda$ to get the effective action $\Gamma_{k=0}$ in principle.
However, due to the conditions for stable calculation mentioned above,
solving the flow equation to small $k$ is quite time-consuming 
for some regions of the $(T,\mu)$ plane, such as the
low-temperature region of the hadronic phase.
In such a region, the curvature of $U_k$, i.e., $m^2_\sigma$, can take a negative value, which leads to small $E_\sigma$ for some $\sigma$. This gives large $F$ and $|G|$ (Eqs.(\ref{FQM}) and (\ref{GQM})) as $k$ decreases and the condition (\ref{finalcond}) becomes difficult to satisfy at small $k$. Thus, some infrared scale $k=k_{\mathrm{IR}}$ is introduced in practice, at which the numerical procedure is stopped. Of course, $k_{\mathrm{IR}}$ should be as small as possible so that sufficiently low-momentum fluctuations are taken into account
to describe the system around the critical point where vanishingly low-momentum excitations exist. Thus we choose a much lower value of $k_{\mathrm{IR}}$ than the $40\,\MeV$ adopted in Ref. \cite{Tripolt:2014wra}, and set $k_{\mathrm{IR}}=1\,\MeV$ as being small enough to incorporate the low-momentum fluctuations.
Therefore our calculation will be reliable
in the vicinity of the critical point,
except for the small surrounding region where
excitation modes with momentum scales lower than $1\,\MeV$
are strongly developed.
Although $\epsilon$, which appears after the analytic
continuation, is defined as a positive infinitesimal,
we set it to $1\,\MeV$ in the present calculation, which should be small enough for present purposes.

3D momentum integrals remain
after the Matsubara summation in Eq. (\ref{G2Sflow}) and Eq. (\ref{G2Pflow}).
However, the integrals can be fully calculated analytically 
for zero external momentum.
Even for a finite external momentum,
they can be nicely reduced to 1D integrals, which are evaluated numerically.
The numerical integrations involve a tricky point, and one has to take care of
the poles of each term in the integrands. 
For example, the first and third terms of the integrand of the second integral in Eq. (\ref{Jabform})
have the same pole $\tilde{E}_{\alpha}=E_{\alpha}+ip_{0}$.
If such terms are integrated separately, a large cancellation
can occur, which then leads to big numerical errors.
Therefore, 
we first combine such  terms analytically in the integrand
before numerical integrations.

\section{Numerical results} \label{SNumericalResults}
\subsection{Parameter setting}

The truncated EAA Eq. (\ref{QMtranc}) and the initial condition
Eq. (\ref{initial}) have some parameters that are fixed so as to
reproduce the observables in vacuum:
We use the same values for the parameters as those in
Ref. \cite{Tripolt:2014wra} and list them in Table \ref{Parameters}.

\begin{table}[!tb]
\begin{center}
\begin{tabular}{ccccc}
\toprule
$\Lambda$ & $m_{\Lambda}/\Lambda$ &
$\lambda_{\Lambda}$ & $c/\Lambda^{3}$ &
$g_{s}$ \\
\hline
1000\,$\mathrm{MeV}$ & 0.794 & 2.00 & 0.001\,75 & 3.2
\\
\bottomrule
\end{tabular}
\end{center}
\caption{Numerical values of the UV scale $\Lambda$ and the parameters in the initial condition
$\Gamma_{k=\Lambda}$ used in the calculation.}
\label{Parameters}
\end{table}
The chiral condensate $\sigma_{0}$ is determined as $\sigma$, which
minimizes $U_{k}(\sigma^{2})-c\sigma$, and the constituent quark mass
$M_{\psi}$ and the sigma and pion screening masses $M_{\sigma}$ and
 $M_{\pi}$ are calculated using Eq. (\ref{Masses}):
\begin{equation}
M_{a}=\left( \left. m^2_{a} \right|_{\sigma=\sigma_{0}, k=k_{\mathrm{IR}}}\right)^{\frac{1}{2}},
\quad \ \ \ (a=\psi, \sigma, \pi).  
\end{equation}
Our parameters reproduce
$\sigma_{0}=93\,\mathrm{MeV}$, $M_{q}=286\,\mathrm{MeV}$, $M_{\pi}=137\,\mathrm{MeV}$, and
$M_{\sigma}=496\,\mathrm{MeV}$ in vacuum.

\subsection{Phase diagram and quark-number susceptibility}
\label{SSPhaseDiagram}

\begin{figure}[!tb]
\centering\includegraphics[width=0.95\columnwidth]{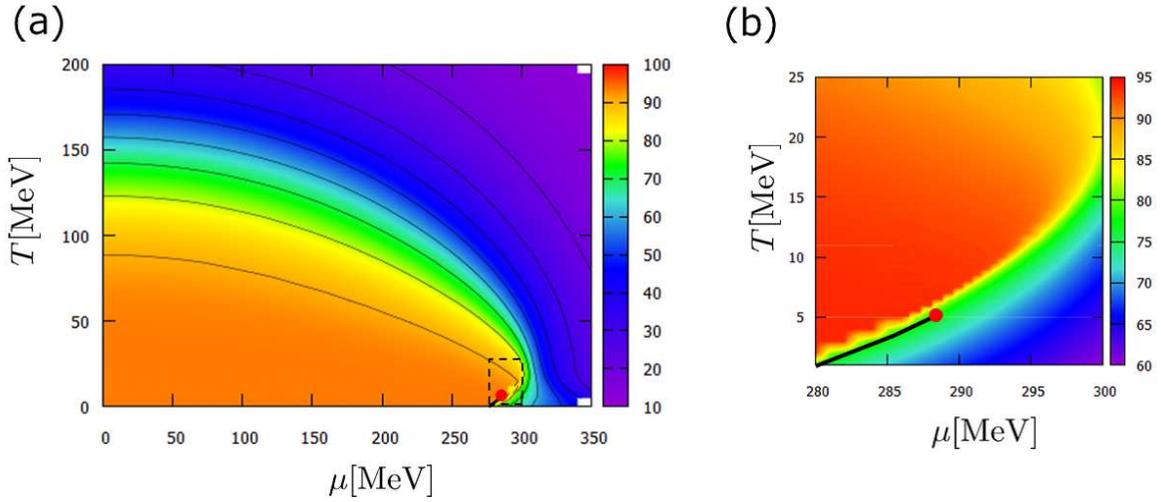}
\caption{(a) Contour map of the chiral condensate in a wide
region. (b) Enlargement of the region surrounded
by dotted lines in (a). The filled circle (red) is the critical point 
and the solid (black) line is the first-order phase boundary.}
\label{phase}
\end{figure}

We show the phase diagram in Fig. \ref{phase}, where a contour map of the chiral condensate is also given.
One sees that chiral restoration occurs as the temperature is raised,
and the phase transition is not a genuine one but a crossover, except for  
the low-temperature and large chemical potential region, where the phase 
transition is of first order. This feature is qualitatively in accordance with 
the results given in the literature, although the location of the critical point here is in a somewhat smaller temperature region than that given in Ref. \cite{Tripolt:2014wra}.
The detailed procedure for locating the critical point is described below.

\begin{figure}[!tb]
\centering\includegraphics[width=0.9\columnwidth]{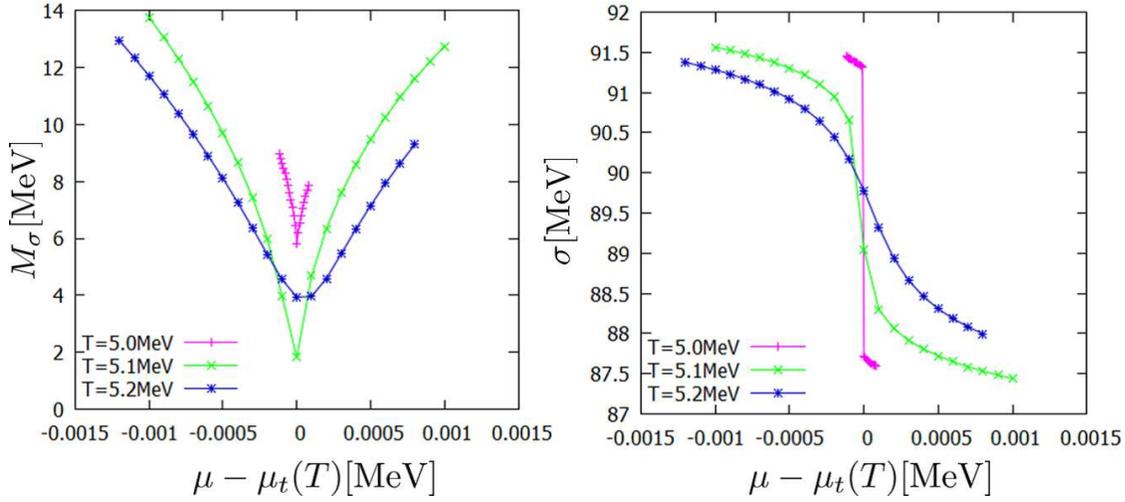}
\caption{The sigma curvature masses and the chiral condensates
at $T=5.0\,\MeV$, $5.1\,\MeV$, and $5.2\,\MeV$
as functions of $\mu-\mu_{t}(T)$.}
\label{Ms}
\end{figure}

At the QCD critical point, the chiral susceptibility diverges.
Therefore, we locate the critical point by searching for the point where the
sigma screening mass $M_{\sigma}$, the square of which is the inverse of the chiral susceptibility,
becomes the smallest: We seek the minimum position of $M_{\sigma}$ using the data points where
$M_{\sigma}$ is greater than $1\,\MeV$,
because our choice of $k_{\mathrm{IR}}=1\,\MeV$ enables us
to take into account fluctuations whose momentum scales are 
greater than $k_{\mathrm{IR}}$ so as to make the result of $M_{\sigma}$ reliable when $M_{\sigma}$ is larger than $1\,\MeV$.
We also identify the first-order phase transition by
a discontinuity of the chiral condensate. The results at
$T=5.0\,\mathrm{MeV}$, $5.1\,\mathrm{MeV}$, and $5.2\,\mathrm{MeV}$
are shown in Fig. \ref{Ms} as functions of $\mu-\mu_{t} (T)$, 
where $\mu_{t} (T)$ is the transition chemical potential
for each temperature determined by the minimum point of the sigma curvature mass and is found to be
$\mu_{t}(5.0\,\MeV)=286.517~02\,\MeV$,
$\mu_{t}(5.1\,\MeV)=286.686~00\,\MeV$, and
$\mu_{t}(5.2\,\MeV)=286.853~20\,\MeV$.

We find that the sigma screening mass becomes smallest between
$T=5.0\,\MeV$ and $T=5.2\,\MeV$ and between $\mu=\mu_{t}(5.0\,\MeV)$ and
$\mu=\mu_{t}(5.2\,\MeV)$. Therefore, the critical temperature $T_{c}$ and
the critical chemical potential $\mu_{c}$ are estimated as 
$T_{c}=5.1\pm 0.1\,\MeV$ and $\mu_{c}=286.6 \pm 0.2\,\MeV$.
The position of the critical point is quite different from  the $(T,\mu)=(10\,\MeV, 292.97\,\MeV)$ given in Ref. \cite{Tripolt:2014wra}.
Such a difference may be attributed to the different choice of
$k_{\mathrm{IR}}$.
In the following discussion,
we regard $T_{c}$ and $\mu_c$ as $5.1\,\MeV$ and
 $286.686\,\MeV$, respectively.
As seen in the behavior of the chiral condensate shown in the
right panel of Fig. \ref{Ms}, the
phase transition along the chemical potential is of first order when $T=5.0\,\MeV$ and a crossover when $T=5.2\,\MeV$.

It is known \cite{Kunhiro:1991} that the quark-number susceptibility  
$\chi_q\equiv\partial \rho_q/\partial \mu$
 is coupled to the scalar susceptibility at finite $\mu$ and can be used
to reveal the nature of the phase transition, where $\rho_q$ denotes the quark-number density.
Indeed $\chi_q$ shows a singular behavior at the critical point, 
and hence an enhancement of {$\chi_q$} is a useful measure of the critical region.
We thus calculate $\chi_q$ to map out the critical region, as was done in Refs.
\cite{Hatta:2002sj,Kamikado:2012cp}.

\begin{figure}[!t]
\centering\includegraphics[width=0.5\columnwidth]{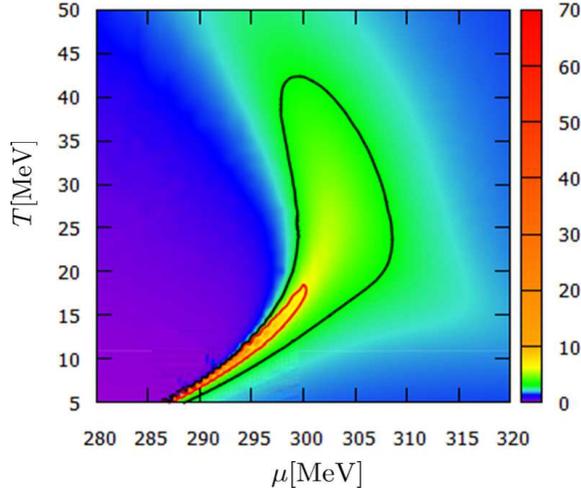}
\caption{Contour map of the quark-number susceptibility near the
 critical point. The values are
 normalized with those of a free quark gas. The bold (black) line is the contour
 line of value 3 and the thin (red) line is that of value 6.}
\label{QNS}
\end{figure}
$\chi_q$ in the FRG formalism is calculated as follows. The effective
action $\Gamma [\pbar,\psi,\sigma,\vec{\pi}]$ in the finite-temperature
formalism is related to the thermodynamic potential $\Omega$ as 
$T\Gamma [0,0,\sigma_{0}, \vec{0}] = \Omega$
under the assumption that the chiral condensate is homogeneous and
$\langle \vec{\pi} \rangle =\vec{0}$,
which in our case reads 
\begin{equation}
\frac{\Omega}{V}= U_{k\rightarrow 0} (\sigma_{0}^{2} ) -c\sigma_{0}.
\label{OmegaU}
\end{equation}
The quark-number susceptibility $\chi_{q}$ is given by differentiating
Eq. (\ref{OmegaU}) twice with respect to the chemical potential:
\begin{equation}
\chi_{q}=-\frac{\partial^{2}}{\partial \mu^{2}}\left(
U_{k\rightarrow 0}(\sigma_{0}^{2})-c\sigma_{0}
\right).
\end{equation}
We carry out the derivatives numerically.
Figure \ref{QNS} shows the contour map of the quark-number susceptibility, normalized by
the value for the massless free quark gas:
\begin{equation}
\chi_{q}=\frac{2N_{c}N_{f}}{6}
\left[ T^{2}+\frac{3\mu^{2}}{\pi^{2}} \right].
\end{equation}

\subsection{Spectral function in the $\sigma$ channel away from the critical point} 
%\sout{at} the hadronic phase and the QGP phases}
\label{SSSpectralexamp}

Before entering into discussions on the spectral properties in the meson channel near the critical point,
we first show the numerical result of the
spectral function $\rho_{\sigma}(\omega, p)$ in the sigma channel
away from the critical point 
in the hadronic and  QGP phases so that the peculiar behavior of the
spectral functions near the critical point shown in the next subsection shall be prominent.

\begin{figure}[!t]
\centering\includegraphics[width=0.9\columnwidth]{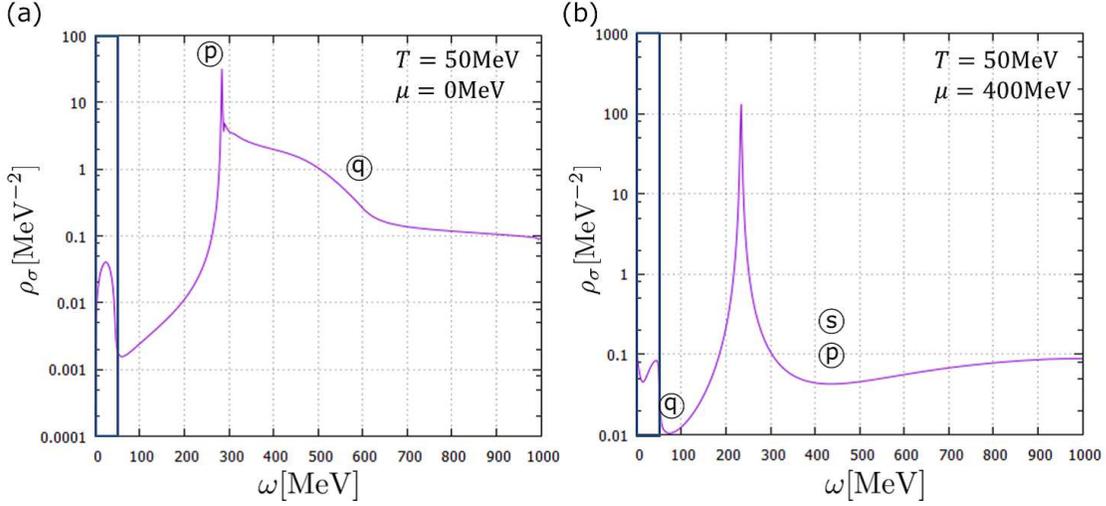}
\caption{
(a) The  spectral function in the $\sigma$ channel  in the hadronic phase at 
($T=50\,\MeV$,\, $\mu=0\,\MeV$).  (b) The same as (a)  in the QGP phase at
($T=50\,\MeV$,\, $\mu=400\,\MeV$).
The spatial momentum is set to $\pab=50\,\MeV$, and the inside of the blue box in each figure is the space-like region ($\omega\,<\,p$).
% are space-like regions.
The positions of the thresholds for the $2\sigma$, $2\pi$, and $\pbar \psi$ decay channels determined by Eq. (\ref{threshold})
are denoted by \textcircled{s}, \textcircled{p}, and \textcircled{q}.
}
\label{SpectP50_1}
\end{figure}
Figures \ref{SpectP50_1}(a) and \ref{SpectP50_1}(b) show
$\rho_{\sigma}(\omega, p)$ at $p\equiv |\vec{p}|=50$ MeV for ($T=50$
MeV, $\mu=0$ MeV) and ($T=50$ MeV, $\mu=400$ MeV), respectively:
The former (latter) is in the hadronic (QGP) phase. 
In the former case, there is a sharp peak at $\omega =
290\,\MeV$ and a relatively small bump in the space-like region $\omega < \pab$;
these correspond to the sigma meson with a modified mass at finite
temperature and the phonon mode composed of particle--hole excitations,
respectively, which is in accord with the result in the RPA in Ref. \cite{Fujii:2004jt}.

The spectral function also tells us the decay and absorption processes
of the particle excitations from the width of the corresponding  peaks or bumps.
In our energy scale, the following processes contribute to the spectral function $\rho_{\sigma}(\omega,\vec{p})$:
\[
\sigma^{\ast}\rightarrow \sigma\sigma, \quad \sigma^{\ast}\rightarrow \pi\pi
, \quad \sigma^{\ast}\rightarrow \pbar \psi, \quad 
\sigma^{\ast}\sigma\rightarrow \sigma,\quad
\sigma^{\ast}\pi\rightarrow \pi, \quad
\sigma^{\ast}\psi \rightarrow \psi,
\]
where $\sigma^{\ast}$ denotes a virtual state in the sigma channel
with energy--momentum $(\omega, \vec{p})$. The energy--momentum
conservation gives constraints on the possible $(\omega, \, \vec{p})$
region for the first three processes as follows:
\begin{align}
\omega \geq \sqrt{\vec{p}^{2}+(2M_{\sigma})^{2}}\ \ &\text{for}\ \ 
\sigma^{\ast}\rightarrow \sigma\sigma, \notag \\
\omega \geq \sqrt{\vec{p}^{2}+(2M_{\pi})^{2}}\ \ &\text{for}\ \ 
\sigma^{\ast}\rightarrow \pi \pi, \label{threshold}\\
\omega \geq \sqrt{\vec{p}^{2}+(2M_{\psi})^{2}}\ \ &\text{for}\ \ 
\sigma^{\ast}\rightarrow \pbar\psi, \notag
\end{align}
which are all in the time-like region. On the other hand, the second
three processes are all collisional ones and possible only in the
space-like region, $0\leq \omega < \pab$.
In particular, the last process $\sigma^{\ast}\psi \rightarrow \psi$
corresponds to the absorption process of the $\sigma^{\ast}$ mode into a
thermally excited quark.
In short, the width of the large bump at $\omega=290\,\MeV$
in Fig.~\ref{SpectP50_1}(a) comes from the $2\pi$ decay process,
while the small bump arises from the space-like processes.

In the latter case, at $T=50\,\MeV$ and $\mu=400\,\MeV$,
the peak position corresponding to the sigma meson is shifted to
$\omega=210\,\MeV$,
while the bump of the particle--hole excitations still persists in the space-like region.

\subsection{Spectral functions near the QCD critical point}
\label{SSSpectral}
\begin{figure}[!t]
\centering\includegraphics[width=0.8\columnwidth]{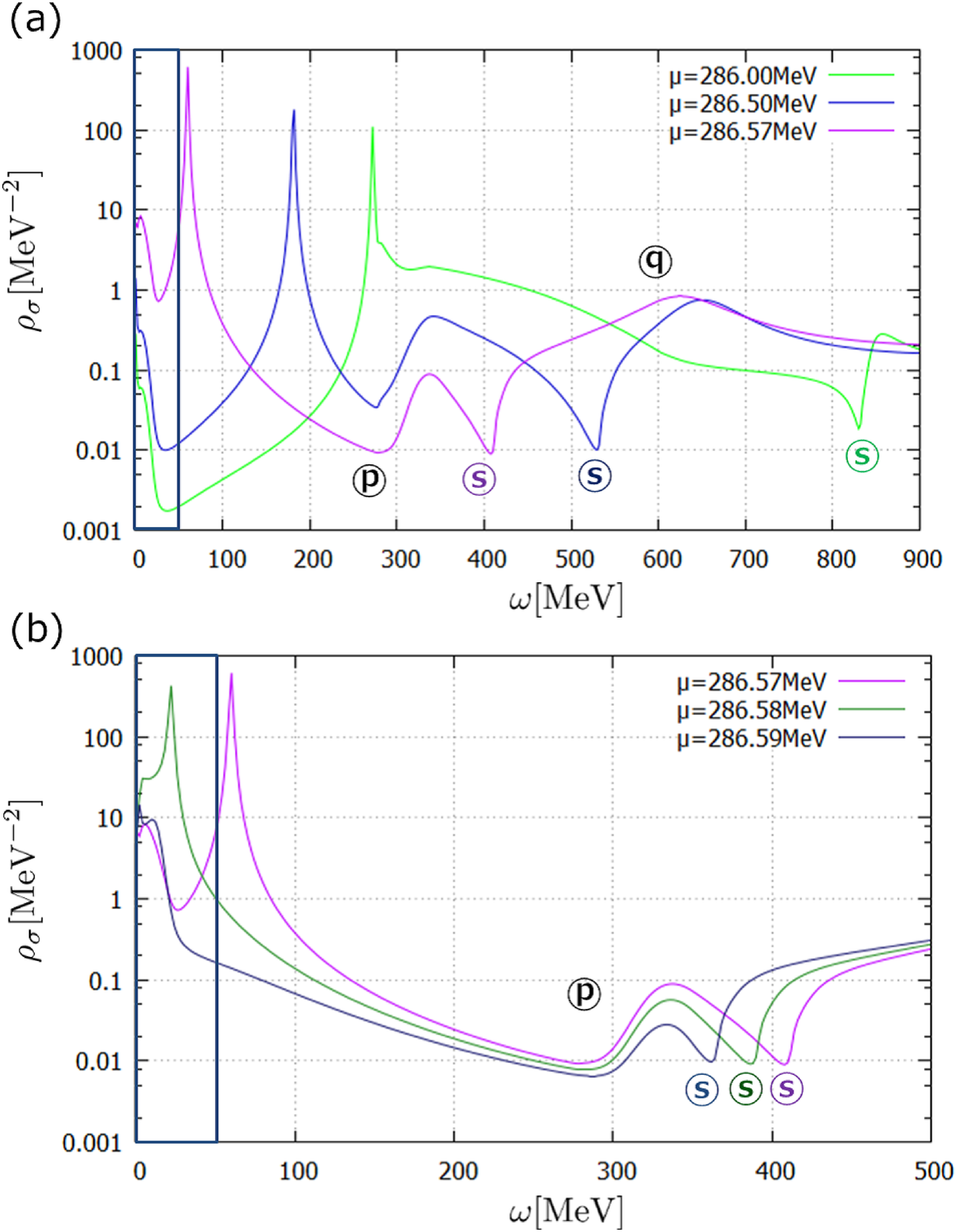}
\caption{The spectral functions in the $\sigma$ channel near the QCD critical point.
The temperature is fixed to $T_{c}$.
The spatial momentum is set to $\pab=50\,\MeV$ and the inside of the blue box in each figure is the space-like region.
The results in $286.00\,\MeV \leq \mu \leq 286.57\,\MeV$ are shown in (a) and
the results in $286.57\,\MeV \leq \mu \leq 286.59\,\MeV$ are shown in (b).
The position of the $2\sigma$ decay threshold for each chemical potential
is denoted by \textcircled{s}.
The $2\pi$ and $\pbar \psi$ decay thresholds hardly change
and are represented by \textcircled{p} and \textcircled{q}.
}
\label{SpectP50_2}
\end{figure}
We calculate the spectral function in the $\sigma$ channel near the QCD
critical point, by increasing the chemical potential toward $\mu_{c}$
along a constant temperature line $T=T_{c}$.
The results at $\mu=286.00\,\MeV$, $\mu=286.50\,\MeV$, and
$\mu=286.57\,\MeV$ are shown in Fig. \ref{SpectP50_2} (a).
One can see the sigma-mesonic peak as well as bumps corresponding
to $2\sigma$ and $2\pi$ decay in the time-like region.
The peak position of the sigma-mesonic mode shifts to lower energy
as the system approaches the critical point.
The position of the $2\sigma$ threshold also shifts to a lower energy
while those of the $2\pi$ and $\pbar \psi$ thresholds hardly
change. The spectral function in the space-like region is drastically
enhanced as the system is close to the critical point.
This behavior can be interpreted as the softening of the
particle--hole mode, which is in accordance with the result
in Ref. \cite{Fujii:2004jt}.
In Fig.~\ref{SpectP50_2}(b), we show the results at chemical potentials much
closer to the critical point. Because of numerical instability 
in $286.60\,\MeV \leq \mu \lesssim 360\,\MeV$,
we choose $\mu =286.58\,\MeV$ and
$\mu =286.59\,\MeV$. 
For comparison, the result at $\mu=286.57\,\MeV$ is also shown.
These results are drastically different from those in 
$\mu \leq 286.57\,\MeV$. In $\mu >286.57\,\MeV$, the peak of the
sigma-mesonic mode penetrates into the space-like region and then
merges into the particle--hole mode.

\begin{figure}[!t]
\centering\includegraphics[width=0.9\columnwidth]{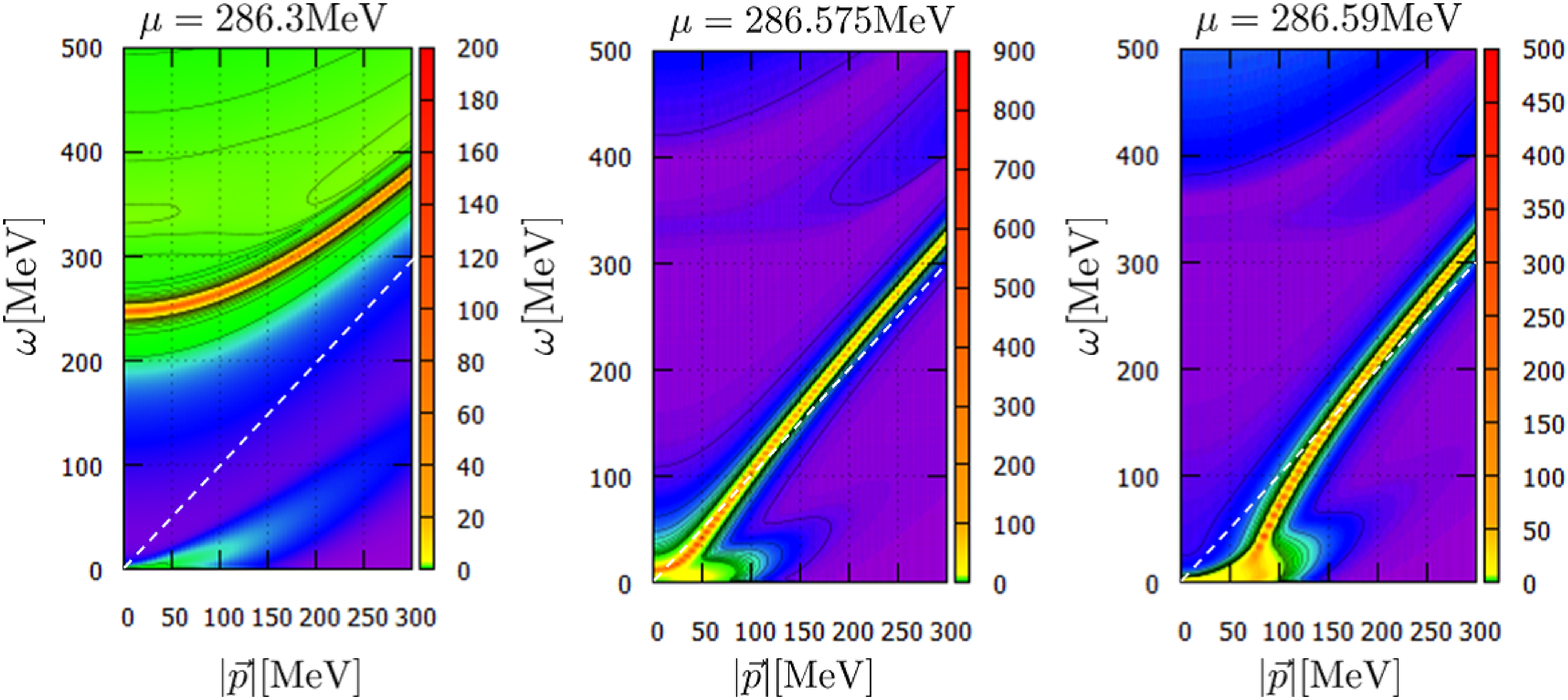}
\caption{Contour maps of $\rho_{\sigma}$
at $T=T_{c}$ and $\mu=286.3\MeV$, $286.575\MeV$, and
$286.59\MeV$.}
\label{Spect3D}
\end{figure}

We can see the dispersion relations of the modes by making contour maps
of the spectral functions as functions of $\omega$ and $\pab$.
Figure \ref{Spect3D} shows the dispersion relations of the sigma-meson and
particle--hole modes near the critical point.
At $\mu=286.3\,\MeV$, the sigma-mesonic peaks can be seen in the time-like
region as well as the particle--hole bump in the space-like region.
As the chemical potential increases, the dispersion relation
of the sigma-mesonic mode shifts downward and it touches  the light cone
near $\mu=286.575\,\MeV$. At $\mu=286.59\,\MeV$, in the low-momentum region
the sigma-mesonic mode clearly penetrates into the space-like region and
merges with the particle--hole bump, which has a  flat dispersion
relation in the low-momentum region.
Our results indicate that the sigma-mesonic mode as well as the
particle--hole mode can become soft near the critical point.

\begin{figure}[!t]
\centering\includegraphics[width=0.95\columnwidth]{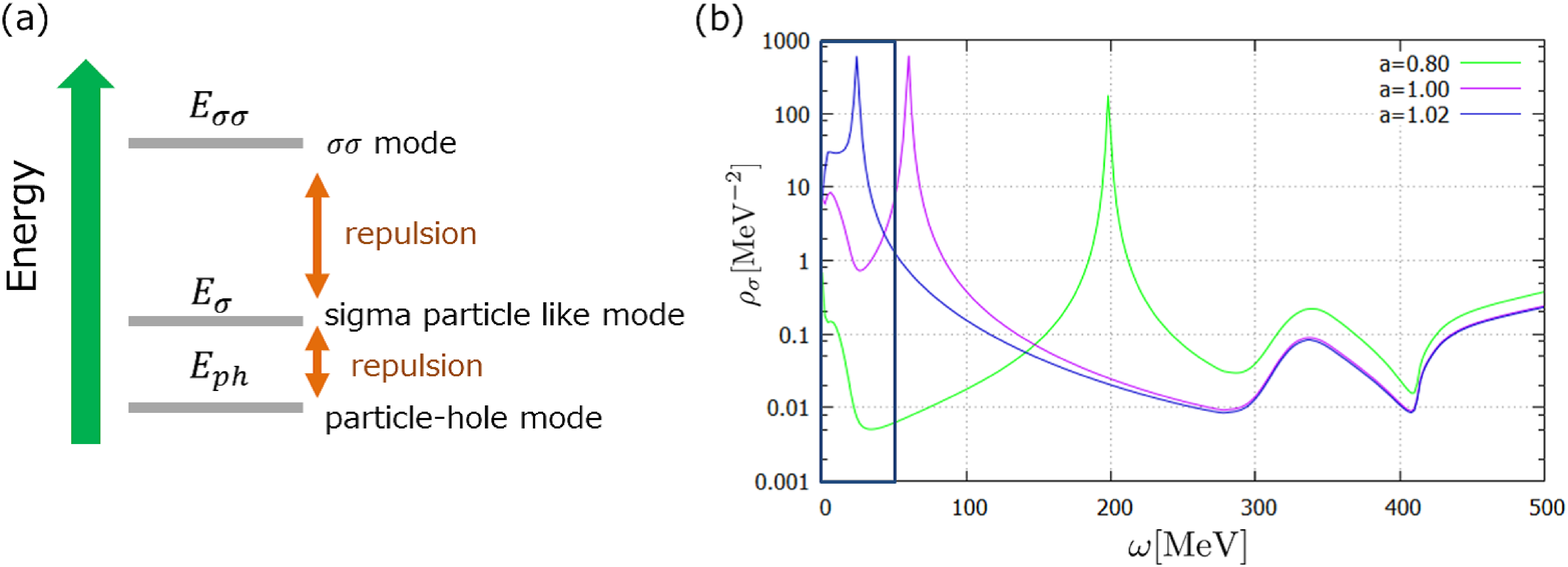}
\caption{(a) Conceptual picture of level repulsion of the sigma-mesonic mode, $2\sigma$ mode, and particle--hole mode. (b) The spectral functions
 in the $\sigma$ channel with a substituted three-point vertex
 $a\Gamma_{k,\sigma\sigma\sigma}^{(3)}$ at $T=T_{c}$,
 $\mu=286.57\,\mathrm{MeV}$, and $\pab=50\,\mathrm{MeV}$. The inside of the blue box is the space-like region.}
\label{G3change}
\end{figure}

One of the possible triggers of this phenomenon is the level
repulsion between the sigma-mesonic mode and other modes.
In particular, the two-sigma ($\sigma\sigma$) mode is considered to play an important role in
the level repulsion since the threshold of the two-sigma mode shifts
downward as the system approaches the critical point.
Let us suppose that the particle--hole mode, the sigma-mesonic mode, and the two-sigma mode can each be described by a state having a single energy level.
Then the system can be regarded as a three-level system,
as depicted in Fig. \ref{G3change} (a):
The interaction within the three states leads to a level repulsion:
If the interaction between the $\sigma$-mesonic mode and  the
$\sigma\sigma$ state becomes sufficiently strong as the system
approaches the critical point,
the energy level of the sigma meson will be so strongly pushed down
that it penetrates into the space-like region.
To show that this scenario can be the case,
we change the strength of the three-point vertex
$\Gamma_{k,\sigma\sigma\sigma}^{(0,3)}$
by hand to investigate the behavior of the sigma-meson peak.
The results in the cases of multiplying
$\Gamma_{k,\sigma\sigma\sigma}^{(0,3)}$ by factors $0.8$ and $1.02$
are shown in Fig. \ref{G3change} (b). The position of the sigma meson
goes up when the three-point vertex is weakened, whereas it exhibits a
downward shift to a lower energy when the three-point vertex is slightly
enhanced. This result suggests that the above interpretation in terms of a level repulsion can be correct.

Here it should be noted that our results exhibit a superluminal group velocity of
the sigma-mesonic mode near the critical point, as seen in
Fig.~\ref{Spect3D} for $\pab=100\,\MeV$ at $\mu=286.59\,\MeV$.
Such an unphysical extreme behavior may 
be an artifact of our truncation scheme, in which some of the higher-order terms in 
the derivative expansion, such as the wave-function renormalization and so on, are neglected,
although a drastic softening of the sigma-mesonic mode may be true. Conversely speaking, such a drawback could disappear
if one uses improved methods with higher-derivative terms being
incorporated. One of the most important improvements 
is the inclusion of
wave-function renormalization
\cite{Helmboldt:2014iya,Kamikado:2013pya},
since this may become important when
additional modes emerge.
However, this task is beyond the scope of the present work and will be
left as a future project.

\begin{figure}[!t]
\centering\includegraphics[width=0.9\columnwidth]{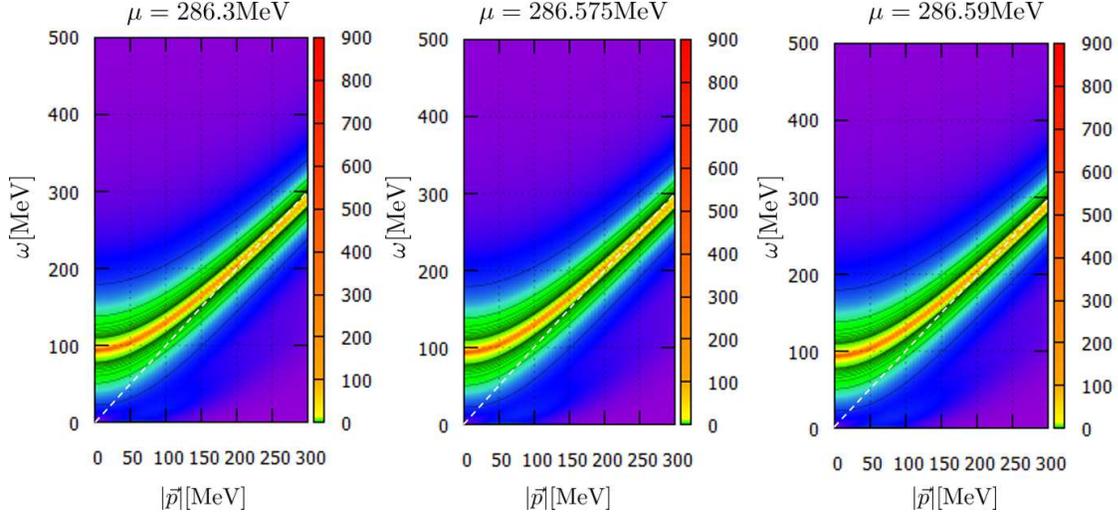}
\caption{Contour maps of $\rho_{\pi}$
at $T=T_{c}$ and $\mu=286.3\,\MeV$, $286.575\,\MeV$, and
$286.59\,\MeV$.}
\label{Spect3DPi}
\end{figure}
So far, we have concentrated on the spectral function in the sigma channel and 
seen interesting behaviors of it near the critical point. It would be intriguing
to examine whether the spectral function $\rho_{\pi}(\omega, p)$ 
in the pion channel shows any peculiar behavior near the critical point.
The numerical result of $\rho_{\pi}(\omega, p)$ near the critical point
is shown in Fig. \ref{Spect3DPi}, from which one can clearly see the dispersion
relation of the pion mode in the time-like region but not in the space-like region.
In contrast to $\rho_{\sigma}$,
$\rho_{\pi}$ hardly changes near the critical point, indicating that there is no critical behavior in the isovector pseudoscalar
modes in either the space-like or time-like region.
This different critical behavior in the sigma and pion channels may be understood as follows: First of all, the finite current quark mass makes the would-be chiral transition 
cease to be of second order, and hence
 prevents the quartet mesons composed of 
the sigma meson and pion
from becoming soft modes inherent in the second-order transition, as mentioned in Sect. \ref{Intro}.
However, the finite chemical potential $\mu\not=0$ makes scalar--vector coupling possible
and the critical point can get to exist with the $Z_2$ universality class of the second-order
transition. Thus the soft modes inherent for the second-order transition 
appear due to the very scalar--vector coupling in 
the space-like region, which is primarily composed of particle--hole excitations. In principle, the pion is coupled to fluctuations in the isovector pseudoscalar or axial vector channels, which are reduced to spin--isospin modes in the nonrelativistic limit\cite{Migdal:1978az}.
Our result simply shows that 
such pionic modes do not develop in the space-like region at least around the critical point.
In the scalar channel, the coupling of the single and double sigma modes is so strong that
a level repulsion between them drastically lowers
the energy of the sigma-mesonic mode in the time-like region.
In contrast to the scalar channel,
the coupling of the single pion mode to other modes, such as the mode that consists of one sigma and one pion,
would not be
strong enough to cause the softening behavior of the isovector pseudoscalar mode in the time-like region.

\section{Summary} \label{SSummary}

We have calculated the  spectral functions in the meson channels in the
quark--meson model with the functional renormalization group method based
on the local potential approximation (LPA). A particular emphasis is put
on the behavior of the spectral function in the $\sigma$ channel near
the QCD critical point. Our results show that the particle--hole mode
(phonon) is enhanced near the critical point,
and thus imply that the density fluctuations are soft modes at the
critical point, as was suggested in the RPA using the NJL
model~\cite{Fujii:2004jt}.
In addition, we have found that the low-energy dispersion curve
of the sigma meson penetrates into the space-like region and the mode
merges into the particle--hole mode near the critical point. 
This result may imply that the sigma meson also acts as a soft mode
at the QCD critical point.
We have also suggested that a possible level repulsion between the
sigma meson and the two-sigma state leads to the anomalous softening of
the sigma meson: An artificial  variation of the strength of the
$\sigma$  three-point vertex $\Gamma_{k,\sigma\sigma\sigma}^{(0,3)}$
strongly affects the position of the sigma meson.
We have also investigated the spectral function in the 
pion channel near the critical point, 
which shows no softening 
in either the time-like region or the space-like region, in contrast to the isoscalar modes. 

Since our result might provide a new picture in which the critical
dynamics  at the QCD critical point can be described by an effective
theory composed of not only the hydrodynamical modes including the
density fluctuation but also the sigma-meson mode, there might be
some implications for the dynamical class of the QCD critical point
\cite{Son:2004iv,Stephanov:1999zu, Berdnikov:1999ph}.

In recent years, the possible existence of inhomogeneous chiral phases in dense quark matter
has been intensively examined \cite{Nakano:2004cd,Nickel:2009wj,Nickel:2009ke,Muller:2013tya}.
It would be interesting to 
investigate in the FRG how the existence of the
inhomogeneous phases affects the behavior of low-energy modes, including the Nambu--Goldstone modes.
It is also worth emphasizing that our analysis of the spectral functions
 in the space-like region or particle--hole modes
can be extended to that of
 precursors of such inhomogeneous phases:
 our results showing the softening and nonsoftening
of the particle--hole modes in the sigma and pion channels might
imply that the inhomogeneous phase with pion condensate 
does not come into existence as a result of a second-order transition.

Our calculation is based on the local potential approximation,
in which some of the higher-order terms in 
the derivative expansion are neglected.
 Such a simple truncation scheme might be an origin of a superluminal group velocity in the close
vicinity of the critical point encountered in Sect. 3.
Thus, it is imperative to 
confirm the results
by employing improved methods incorporating higher-derivative terms, including the 
wave-function renormalization, since it is important to identify when
composite collective modes emerge.
This intriguing task will, however,
be left for future work.

\section*{Acknowledgements}
T.~K.~is supported by JSPS KAKENHI Grants No. 24340054
and  by the Yukawa International Program for Quark--Hadron Sciences (YIPQS).
K.~M.~was supported by Grants-in-Aid for Scientific Research on
Innovative Areas from MEXT (Grant No. 24105008).
The numerical computation in this work was carried out at the 
Yukawa Institute Computer Facility.

\appendix
\section{The explicit forms of loop-functions after Matsubara summations } \label{AppFlow}
In this appendix, we show the explicit forms of Eqs. (\ref{Jab})--(\ref{Jqq}) after Matsubara summation. The momentum integral of $I^{(2)}_{\alpha}$
can be calculated easily and its form is given as
follows:
\begin{equation}
I^{(2)}_{\alpha}=\frac{k^4}{6\pi^2}\left(
\frac{1+2n_{B}(E_{\alpha})}{2E_{\alpha}^{3}}-\frac{n_{B}^{\prime}(E_{\alpha})}{E_{\alpha}^2}\right),
\end{equation}
where $E_{\alpha}=\sqrt{k^{2}+m_{\alpha}^{2}}$ and $n_{B,F}^{\prime}(E)=\frac{dn_{B,F}(E)}{dE}$.

Next, we show the forms of $J_{k,\alpha \alpha}(\pfour)$
and $J_{k,\pbar \psi}^{(\sigma)}(\pfour)$ after Matsubara summation.
$J_{k,\alpha \alpha}(\pfour)$ ($J_{k,\pbar \psi}^{(\sigma)}(\pfour)$)
has regulators with different arguments: $R_{k}^{B}(\qfour)$ and $R_{k}^{B}(\qfour-\pfour)$
($R_{k}^{F}(\qfour)$ and $R_{k}^{F}(\qfour-\pfour)$). These regulators
contain the Heaviside step functions such that  the momentum dependence of the integrands differs
between the two integral regions $D_{1}$ and $D_{2}$:
\begin{align}
D_1&=\left\lbrace
\left. \vec{q} \in \mathbb{R}^{3} \right|
|\vec{q}-\vec{p}|<k\ \text{and}\ |\vec{q}|<k 
\right\rbrace, \notag
\\ 
D_2&=\left\lbrace
\left. \vec{q} \in \mathbb{R}^{3} \right|
|\vec{q}-\vec{p}|<k\ \text{and}\ |\vec{q}|>k 
\right\rbrace.
\notag
\end{align}

Using the following notations,
\begin{align}
\tilde{E}_{\alpha}=\sqrt{\vec{q}^2+M_{\alpha}^{2}},\ \ \ \ 
\cos \varphi=\frac{\vec{q}\cdot (\vec{q}-\vec{p})}
{\left|\vec{q} \right| \left|\vec{q}-\vec{p}\right|},
\notag
\end{align}
we show the forms of $J_{k,\alpha \alpha}(\pfour)$
and $\Jsqq(\pfour)$ after Matsubara summation below:
\begin{align}
\Jab(\pfour)
=&\int_{D_{1}} \frac{d^3 q}{\left (2\pi\right )^{3}}
\frac{k}{2}
\left[
(1+n_B(\Ea))\frac{\Ea^2+\Eb^2-(2\Ea+\ip0)^2}{\Ea^3(\Eb^2-(\Ea+\ip0)^2)^2}
+n_B(\Ea)\frac{\Ea^2+\Eb^2-(2\Ea-\ip0)^2}{\Ea^3(\Eb^2-(\Ea-\ip0)^2)^2}  \right. \notag \\
&+\frac{2(1+n_B(\Eb))}{\Eb(\Ea^2-(\Eb-\ip0)^2)^2}
+\frac{2n_B(\Eb)}{\Eb(\Ea^2-(\Eb+\ip0)^2)^2}
\notag \\
&\left. -\frac{n_{B}^{\prime}(\Ea)}{\Ea^2(\Eb^2-(\Ea-\ip0)^2)}
-\frac{n_{B}^{\prime}(\Ea)}{\Ea^2(\Eb^2-(\Ea+\ip0)^2)} \right] \notag \\
&+\int_{D_{2}} \frac{d^3 q}{\left (2\pi\right )^{3}}
\frac{k}{2}\left[(1+n_B(\Ea))\frac{\Ea^2+\Ebt^2-(2\Ea+\ip0)^2}{\Ea^3(\Ebt^2-(\Ea+\ip0)^2)^2}
+n_B(\Ea)\frac{\Ea^2+\Ebt^2-(2\Ea-\ip0)^2}{\Ea^3(\Ebt^2-(\Ea-\ip0)^2)^2} \right. \notag \\
&+\frac{2(1+n_B(\Ebt))}{\Ebt(\Ea^2-(\Ebt-\ip0)^2)^2}
+\frac{2n_B(\Ebt)}{\Ebt(\Ea^2-(\Ebt+\ip0)^2)^2}
\notag \\
&\left. -\frac{n_{B}^{\prime}(\Ea)}{\Ea^2(\Ebt^2-(\Ea-\ip0)^2)}
-\frac{n_{B}^{\prime}(\Ea)}{\Ea^2(\Ebt^2-(\Ea+\ip0)^2)} \right],
\label{Jabform}
\end{align}
\begin{align}
\Jsqq(\pfour)=&\Jpqq(\pfour)+\int_{D_1}\frac{d^3 q}{(2\pi)^3}
4m_{\psi}^2 g_{s}^2k\left[
(1-n_F(\Eq-\mu)-n_F(\Eq+\mu))\frac{12\Eq^2+\p02}{\Eq^3(4\Eq^2+\p02)^2} \right. \notag \\
&\left. -n_F^{\prime}(\Eq-\mu)\frac{1}{\ip0\Eq^2(2\Eq+\ip0)}
+n_F^{\prime}(\Eq+\mu)\frac{1}{\ip0\Eq^2(2\Eq-\ip0)}
\right] \notag
\\
&+\int_{D_2}\frac{d^3 q}{(2\pi)^3}
4m_{\psi}^2 g_{s}^2 k \left[
(1-n_F(\Eq-\mu))\frac{(2\Eq-\ip0)^2-6\Eq^2-k^2+2\p02+\q2}{\Eq^3(\Eqt^2-(\Eq+\ip0)^2)^2}\right.
\notag \\
&-n_F(\Eq+\mu)\frac{(2\Eq+\ip0)^2-6\Eq^2-k^2+2\p02+\q2}{\Eq^3(\Eqt^2-(\Eq-\ip0)^2)^2} \notag \\
&+\frac{n_F^{\prime}(\Eq-\mu)}{\Eq^2(\Eqt^2-(\Eq+\ip0)^2)}
+\frac{n_F^{\prime}(\Eq+\mu)}{\Eq^2(\Eqt^2-(\Eq-\ip0)^2)} \notag
\\
&\left.+\frac{2(1-n_F(\Eqt-\mu))}{\Eqt(\Eq^2-(\Eqt-\ip0)^2)^2}
-\frac{2n_F(\Eqt+\mu)}{\Eqt(\Eq^2-(\Eqt+\ip0)^2)^2} \right],
\label{Jsform}
\end{align}
\begin{align}
\Jpqq(\pfour)=&
\int_{D_1}\frac{d^3 q}{(2\pi)^3}
\left (-2g_s^2k\right ) \left[(1-n_F(\Eq-\mu)-n_F(\Eq+\mu))\frac{16\Eq^4-k^2(12\Eq^2+\p02)}{\Eq^3(4\Eq^2+\p02)^2} \right. \notag \\
&\left. +n_F^{\prime}(\Eq-\mu)\frac{k^2+\ip0\Eq}{\ip0\Eq^2(2\Eq+\ip0)}
-n_F^{\prime}(\Eq+\mu)\frac{k^2-\ip0\Eq}{\ip0\Eq^2(2\Eq-\ip0)} \right] \notag \\
&+\int_{D_1}\frac{d^3 q}{(2\pi)^3}\cos\varphi
\left (-2g_s^2k\right ) \left[(1-n_F(\Eq-\mu)-n_F(\Eq+\mu))\frac{-8\Eq^4+2\Eq^2(6k^2-\p02)+k^2\p02}{\Eq^3(4\Eq^2+\p02)^2} \right. \notag \\
&\left. -\frac{k^2n_F^{\prime}(\Eq-\mu)}{\ip0\Eq^2(2\Eq+\ip0)}
+\frac{k^2n_F^{\prime}(\Eq+\mu)}{\ip0\Eq^2(2\Eq-\ip0)} \right] \notag \\
&+\int_{D_2}\frac{d^3 q}{(2\pi)^3}
\left (-2g_s^2k\right ) \left[(1-n_F(\Eq-\mu))
\frac{-2\ip0\Eq^3+4\ip0 k^2 \Eq-k^2(\p02-k^2+\q2)+2\q2\Eq^2}{\Eq^3 (\Eqt^2-(\Eq+\ip0)^2)^2}
\right. \notag \\
&+n_F(\Eq+\mu)
\frac{-2\ip0\Eq^3+4\ip0 k^2 \Eq+k^2(\p02-k^2+\q2)-2\q2\Eq^2}{\Eq^3 (\Eqt^2-(\Eq-\ip0)^2)^2}\notag \\
&-n_F^{\prime}(\Eq-\mu)\frac{k^2+\ip0\Eq}{\Eq^2(\Eqt^2-(\Eq+\ip0)^2)}
-n_F^{\prime}(\Eq+\mu)\frac{k^2-\ip0\Eq}{\Eq^2(\Eqt^2-(\Eq-\ip0)^2)} \notag \\
&\left.-(1-n_F(\Eqt-\mu))\frac{2(\q2-\ip0\Eqt)}{\Eqt(\Eq^2-(\Eqt-\ip0)^2)^2}
+n_F(\Eqt+\mu)\frac{2(\q2+\ip0\Eqt)}{\Eqt(\Eq^2-(\Eqt+\ip0)^2)^2} \right] \notag \\
&+\int_{D_2}\frac{d^3 q}{(2\pi)^3}\cos\varphi
\left (-2g_s^2k\right )\qabs \notag \\
&\times \left[
(1-n_F(\Eq-\mu))
\frac{2\ip0 \Eq^3-4\ip0 \Eq k^2+k^2(\p02-k^2+\q2)-\Eq ^2 (k^2+\p02+\q2)}
{\Eq^3 k(\Eqt^2-(\Eq+\ip0)^2)^2} \right.
\notag \\
&+n_F(\Eq+\mu)
\frac{2\ip0 \Eq^3-4\ip0 \Eq k^2-k^2(\p02-k^2+\q2)+\Eq ^2 (k^2+\p02+\q2)}
{\Eq^3 k(\Eqt^2-(\Eq-\ip0)^2)^2} \notag \\
&+\frac{k n_F^{\prime}(\Eq-\mu)}{\Eq^2(\Eqt^2-(\Eq+\ip0)^2)}
+\frac{k n_F^{\prime}(\Eq+\mu)}{\Eq^2(\Eqt^2-(\Eq-\ip0)^2)} \notag \\
&\left. +(1-n_F(\Eqt-\mu))\frac{k^2-\p02+\q2-2\ip0\Eqt}{k\Eqt(\Eq^2-(\Eqt-\ip0)^2)^2}
-n_F(\Eqt+\mu)\frac{k^2-\p02+\q2+2\ip0\Eqt}{k\Eqt(\Eq^2-(\Eqt+\ip0)^2)^2} \right],
\label{Jpform}
\end{align}
for which the external Matsubara frequency $ip_{0}$ is to
be replaced to make analytic continuation.

\section{Numerical stability conditions for solving the flow equation of the effective potential}
\label{AppStability}
In general when one solves a partial differential equation
numerically, the discretization of derivatives may cause numerical
instability. Thus, one needs to impose numerical stability conditions to 
avoid the enhancement of the error due to accumulation.
The derivation of such conditions is concretely demonstrated
in the case of linear partial differential equations and briefly mentioned in the case of nonlinear partial
differential equations in Ref. \cite{NumericalRecipes}.
In this appendix we consider the numerical stability
conditions for a nonlinear partial differential
equation that is a generalized equation of Eq. (\ref{Ukflow})
to derive the numerical stability conditions for solving Eq. (\ref{Ukflow}) in the grid method.

Consider a nonlinear partial differential equation
for some function $u(t,\sigma)$ of the following form:
\begin{equation}
\frac{\partial u(t,\sigma)}{\partial t}=f\left(t,\sigma,u(t,\sigma), \frac{\partial u(t,\sigma)}{\partial \sigma},\frac{\partial^{2}u(t,\sigma)}{\partial \sigma^{2}}\right),
\label{flow}
\end{equation}
where $f$ is an arbitrary real function.
This equation is a generalized equation of Eq. (\ref{Ukflow}).
We derive the numerical stability conditions for this equation in the case of the forward difference for the $t$-derivative and the central three-point difference for the $\sigma$-derivative.
We expect that the derived conditions can also be applied to the fourth-order
Runge--Kutta method that we use in the practical calculations.
Then the discretized flow equation leads
\begin{align}
&\frac{u(t+\Delta t,\sigma)-u(t,\sigma)}{\Delta t} \notag \\
&=f\left(t,\sigma,u(t,\sigma), \frac{u(t,\sigma+\Delta \sigma)-u(t,\sigma-\Delta \sigma)}
{2\Delta \sigma},\frac{u(t,\sigma+\Delta \sigma)-2u(t,\sigma)
+u(t,\sigma-\Delta \sigma)}{{(\Delta \sigma)}^{2}}\right),
\label{Dflow}
\end{align}
where $\Delta t$ and $\Delta \sigma$ are intervals
of discretization of $t$ and $\sigma$.
We suppose that some numerical error occurs
at some step of the numerical calculation and $u(t,\sigma)$ is written as follows:
\begin{equation}
u(t,\sigma)=u_{0}(t,\sigma)+\delta u(t,\sigma),
\label{uerror}
\end{equation}
where $u_{0}(t,\sigma)$ is the exact solution of Eq. (\ref{Dflow})
and $\delta u(t,\sigma)$ is the deviation from the exact solution
caused by numerical error.
We substitute Eq. (\ref{uerror}) into Eq. (\ref{Dflow})
and get the evolution equation for $\delta u(t,\sigma)$, considering the first order of expansion with $\delta u(t,\sigma)$
and using the fact that $u_{0}(t,\sigma)$ exactly satisfies Eq. (\ref{Dflow}):
\begin{align}
\frac{\delta u (t+\Delta t, \sigma)-\delta u (t,\sigma)}{\Delta t}
&=F\left(t,\sigma,u_{0},\frac{\tilde{\partial} u_{0}}{\partial \sigma},\frac{\tilde{\partial}^{2} u_{0}}{\partial \sigma^{2}}\right)
\frac{\delta u(t+\Delta t,\sigma)-\delta u(t-\Delta t,\sigma)}{2\Delta \sigma}
\notag \\
&+G\left(t,\sigma,u_{0},\frac{\tilde{\partial} u_{0}}{\partial \sigma},\frac{\tilde{\partial}^{2}u_{0}}{\partial \sigma^{2}}\right)\frac{\delta u(t+\Delta t,\sigma)-2\delta u(t,\sigma)+\delta u(t-\Delta t,\sigma)}{{(\Delta \sigma)}^{2}},
\label{flow_increment}
\end{align}
where
\begin{equation}
F\equiv\frac{\partial f}{\partial u'},\ 
G\equiv\frac{\partial f}{\partial u''}
\  \left(u'=\frac{\partial u}{\partial \sigma}, u''= \frac{\partial^{2} u}{\partial \sigma^{2}} \right), \notag
\end{equation}
\begin{equation}
\frac{\tilde{\partial}u_{0}}{\partial \sigma}
\equiv\frac{\delta u(t,\sigma +\Delta \sigma)-\delta u(t,\sigma -\Delta \sigma)}{2\Delta \sigma},
\ 
\frac{\tilde{\partial}^{2}u_{0}}{\partial \sigma^{2}}
\equiv\frac{\delta u(t,\sigma + \Delta \sigma)-2\delta u(t,\sigma)+\delta u(t,\sigma - \Delta \sigma)}{{(\Delta \sigma)}^{2}}. \notag
\end{equation}
For simplicity, we ignore the $\sigma$ dependence of $F$ and $G$.
Then the solutions of Eq. (\ref{flow_increment})
can be written as linear combinations of Fourier components:
\begin{equation}
\delta u(m\Delta t,n \Delta \sigma;k)= \xi_{k}(m\Delta t) e^{ikn\Delta \sigma},
\label{fourier}
\end{equation}
where we substitute $m\Delta t$ and $n\Delta \sigma$ for $t$ and $\sigma$ ($m,n\in \mathbb{Z}$).
By substituting Eq. (\ref{fourier}) into Eq. (\ref{flow_increment}), we get the following equation
after some manipulation:
\begin{equation}
\frac{\xi_{k}((m+1)\Delta t)}{\xi_{k}(m\Delta t)}
=1-\frac{2G\Delta t}{(\Delta \sigma)^{2}}(1-\cos k\Delta \sigma)
+iF\frac{\Delta t}{\Delta \sigma}\sin(k\Delta \sigma).
\label{xi_evolution}
\end{equation}
If $|\xi_{k}((m+1)\Delta t)/\xi_{k}(m\Delta t)|>1$,
the numerical deviation is amplified as $m$ increases (the flow step goes forward) and numerical instability occurs.
Therefore, the condition for stable calculation is
\begin{equation}
\left|\frac{\xi_{k}((m+1)\Delta t)}{\xi_{k}(m\Delta t)} \right|\leq1.
\end{equation}
From Eq. (\ref{xi_evolution}), this condition can be
rewritten as
\begin{equation}
h(X)\leq 1\ \ (-1\leq X \leq 1),
\label{ineq}
\end{equation}
where $X\equiv \cos (k\Delta x)$ and $h(X)$ is defined as
\begin{equation}
h(X)\equiv(a^{2}-b^{2})X^{2}+2a(1-a)X+(1-a)^{2}+b^{2}
\ \ \left(a=\frac{2G\Delta t}{(\Delta \sigma)^{2}}, b=\frac{F\Delta t}{\Delta \sigma} \right).
\end{equation}
Considering that $h(X)$ satisfies $h(1)=1$,
one can easily understand that the condition Eq. (\ref{ineq})
is equivalent to the following conditions:
\begin{numcases}
{}
\frac{dh}{dX}(X=1)\geq 0, & \notag \\
h(-1) \leq 1. &
\end{numcases}
These conditions are rewritten as follows
\begin{numcases}
{}
2(a^{2}-b^{2})+2a(1-a)\geq 0, & \notag \\
a^{2}-b^{2}-2a(1-a)+(1-a)^{2}+b^{2}\leq 1. \label{ineq1}&
\end{numcases}
We show later that $G<0$ and $\Delta t <0$ in the case of
Eq. (\ref{Ukflow}).
Supposing that $G<0$ and $\Delta t <0$, Eq. (\ref{ineq1})
is finally rewritten as
\begin{numcases}
{}
|\Delta t| \leq \frac{2|G|}{F^{2}}, & \notag \\
|\Delta t| \leq \frac{\Delta \sigma^{2}}{2|G|}.& \label{finalcond}
\end{numcases}

In the case of Eq. (\ref{Ukflow}),
$t$ and $u(t,\sigma)$ are identified with $\ln(k/\Lambda)$ and $U_{k}(\sigma^{2})$, respectively,
and $F$ and $G$ are derived to be:
\begin{numcases}
{}
F=-\frac{k^{5}}{8\pi^{2}\sigma E_{\sigma}^{3}}
\left(\coth\frac{E_{\pi}}{2T}+\frac{E_{\pi}}{2T}\frac{1}{\sinh^{2}\frac{E_{\pi}}{2T}}\right), & \label{FQM}\\
G=-\frac{k^{5}}{24\pi^{2}E_{\sigma}^{3}}
\frac{\coth\frac{E_{\sigma}}{2T}+\frac{E_{\sigma}}{2T}\frac{1}{\sinh^{2}\frac{E_{\sigma}}{2T}}}
{\left(\coth\frac{E_{\pi}}{2T}+\frac{E_{\pi}}{2T}\frac{1}{\sinh^{2}\frac{E_{\pi}}{2T}}\right)^{2}}. & \label{GQM}
\end{numcases}
$G$ is negative definite, and $\Delta t$ is also negative
because the direction of flow is from $k=\Lambda$ to $k=0$.

Although the above discussion contains rough approximations,
numerical instability can be avoided by adjusting $\Delta t$
and $\Delta \sigma$ so that Eq. (\ref{finalcond}) is satisfied.
Because the above condition of Eq. (\ref{finalcond}) is
too strict when $\sigma$ is close to zero, we neglect
the condition around $\sigma=0$.

%%%%%%%%%%%%%%%%%%%%%%%%%%%%%%%

\end{document}